\documentclass[twocolumn,showpacs,aps,amssymb,preprintnumbers]{revtex4}

\usepackage{graphicx}
\usepackage{dcolumn}
\usepackage{bm}

\newcommand{\lsim}   {\mathrel{\mathop{\kern 0pt \rlap
  {\raise.2ex\hbox{$<$}}}
  \lower.9ex\hbox{\kern-.190em $\sim$}}}
\newcommand{\gsim}   {\mathrel{\mathop{\kern 0pt \rlap
  {\raise.2ex\hbox{$>$}}}
  \lower.9ex\hbox{\kern-.190em $\sim$}}}
\begin{document}
\preprint{MPP-2004-149}
\title{Gravitational radiation from rotating monopole-string systems}

\author{E.~Babichev$^{1,2}$, V.~Dokuchaev$^2$ and M.~Kachelrie{\ss}$^1$}

\affiliation{
$^1$Max-Planck-Institut f\"ur Physik
(Werner-Heisenberg-Institut), D--80805 M\"unchen, Germany}

\affiliation{
$^2$Institute for Nuclear Research of the Russian Academy of
Sciences, 60th October Anniversary Prospect 7a, 117312 Moscow, Russia}

\date{\today}

\begin{abstract}
We study the gravitational radiation from a rotating monopole-antimonopole
pair connected by a string. While at not too high frequencies the
emitted gravitational spectrum is described asymptotically by
$P_n\propto n^{-1}$,  the spectrum is exponentially suppressed in the
high-frequency limit, $P_n\propto \exp(-n/n_{\rm cr})$. Below $n_{\rm cr}$,
the emitted spectrum of gravitational waves is very similar to the
case of an oscillating monopole pair connected by
a string, and we argue therefore that the spectrum found holds
approximately for any moving monopole-string system.
As application, we discuss the stochastic gravitational wave background
generated by monopole-antimonopole pairs connected by strings
in the early Universe and gravitational wave bursts
emitted at present by monopole-string networks.
We confirm that advanced gravitational wave detectors have
the potential to detect a signal for string tensions as small as
$G\mu\sim 10^{-13}$.
\end{abstract}


\maketitle

\section{Introduction}

The formation of topological defects during symmetry-breaking phase
transitions in the early Universe~\cite{Kibble} is a generic
prediction of Grand Unified Theories. Most prominent example is
the breaking of a semi-simple group $G\to H\otimes U(1)$ giving rise
to monopoles. If in a second phase transition the $U(1)$ is broken
down to $Z_N$,
\begin{equation}
  \label{G1}
   G\to H\otimes U(1)\to H \otimes Z_N \,,
\end{equation}
each monopole gets attached to $N$ strings. For $N>1$, a
long-lived monopole-string network (``necklaces'') forms  which may
evolve towards the  scaling regime with several long (infinite)
strings and many closed string loops inside the Hubble horizon.
The case $N=1$, i.e.
\begin{equation}
  \label{G2}
   G\to H\otimes U(1)\to H \,,
\end{equation}
is rather different: the second phase transition
results in single monopole-antimonopole pairs connected by one string.
Typically, the life-time of this system is much shorter than the
Hubble time and it does not survive until present.

An important signature of both systems is the emitted gravitational
radiation. Oscillations of monopole-antimonopole pairs connected by strings
in the early Universe may form a stochastic gravitational wave
background which is---as shown in Ref.~\cite{MarVil1}---detectable for
advanced gravitational wave detectors like LIGO~\cite{LIGO},
VIRGO~\cite{VIRGO} and LISA~\cite{LISA}. For the case of
string networks  that survive until present, Damour and
Vilenkin~\cite{DamVil1,DamVil2}
considered the gravitational radiation from cusps of
a string network. They argued that the resulting
Gravitational Wave Bursts (GWBs) may be detectable for string tensions
as small as $G\mu\sim 10^{-13}$.

In this work, we investigate the gravitational radiation from a simple
rotating system consisting of a straight string and two monopoles
attached to its ends. We assume that the (anti-) monopole
charges are confined inside the string and thus the main
energy losses of the topological defects is the emission of
gravitational radiation. 
The solution to the equation of motion for such
a system, dubbed ''rotating rod'', was found in Ref.~\cite{MarVil}.
We derive the spectrum of gravitational waves emitted and simple
asymptotic formulas for different frequency ranges.
We find that the dimensionless parameter $\mu R/m$, where $\mu$ is the
string tension, $R$ the distance between monopoles and $m$ the monopole
mass,
determines the main features of the emitted spectrum: For not too high
mode numbers,
$n \ll n_{\rm cr}=(\mu R/m)^{3/2}$, the gravitational wave spectrum
can be approximated by  $P_n\approx 5.77 G\mu^2/n$. In the high-frequency
limit, $n\gg n_{\rm cr}$, the gravitational radiation is
exponentially suppressed.
For moderate mode numbers, $n\ll n_{\rm cr}$, in which most
energy is radiated, the spectrum obtained is very similar to the one
of an oscillating monopole-antimonopole pair connected by a string.
We conclude therefore that the spectrum found applies also to
the generic case of a superposition of a rotational and oscillatory motion
of the monopole-string system.

Using the obtained gravitational radiation spectrum of the rotating rod
solution we study how the presence of monopoles and antimonopoles in a
string network influences the possibility to observe  GWBs. We find
that the presence of monopoles for any mass below $M_{\rm Pl}$ does not
affect the possibility of GWBs detection.

For the case of a symmetry-breaking scheme with $N=1$, we estimate
how the differences in the emission spectra between the rotating and
oscillating  monopole pair connected by string changes the predictions
of Ref.~\cite{MarVil1} for the produced gravitational wave
background. We find that the different high-frequency behavior of the
two solutions only marginally influences the results.

The paper is organized as follows. In Sec.~\ref{GR}, we study the
emission of gravitational radiation from a rotating rod in detail.
Then we compare our results with the spectrum  of another simple
solution considered in Ref.~\cite{MarVil}. In Sec.~\ref{network},
we investigate the potential of gravitational wave detectors to
observe GWBs from a string-monopole network or the gravitational wave
background from $N=1$ monopole-string systems.
We summarize briefly our results in Sec. \ref{Conclusion}.

\section{Gravitational radiation}
\label{GR}
The equations of motion for two monopoles connected by a string of
length $L$ can be solved analytically only in special cases. One case
was dubbed ``rotating rod'' by the authors of Ref.~\cite{MarVil} and
is given by
\begin{eqnarray}
  \label{sol}
  \mathbf{x}(\sigma,t)&=&R\,\sin(\sigma/R)\,\mathbf{y}(t/R),\nonumber\\
  \mathbf{x}_i(t)&=&\pm(L/2)\mathbf{y}(t/R),\\
  \sigma_i&=&\pm R\arcsin(L/2R),\nonumber
\end{eqnarray}
where $\mathbf{x}(\sigma/R)$ is the coordinate of the string at time $t$
and position $\sigma$, $\textbf{x}_i(t)$ are the positions of two
point-like masses ($i=1$, $2$), $\sigma_i$ are the spatial parameters of
the string at the positions of the two monopoles,
$\textbf{y}(\theta)=(\cos\theta,\,\sin(\theta))$,
$L=[(1+4a^2R^2)^{1/2}-1]/a$ and $a=\mu/m$. The system (\ref{sol}) has
the period $T=\pi R$.

As for any periodic system, we can introduce the Fourier transform
of the energy-momentum tensor $T^{\mu\nu}(\mathbf{x},t)$ as
\begin{equation}
  \label{Fou:Tstr}
  \hat{T}^{\mu\nu}(\omega_{n},\mathbf{l})=
  \frac{1}{T}\int_0^T dt\int d^{3}x\: T^{\mu\nu}(\mathbf{x},t)
  e^{i(\omega_{n}t-\mathbf{l}\mathbf{x})} \,,
\end{equation}
where $\omega_{l}=2\pi l/T$ and $\mathbf{l}$ is an arbitrary unit
vector. The following calculations are simplified in the corotating
basis associated with the vector $\mathbf{l}$~\cite{Durrer}: In the
new basis $(\mathbf{e}_{1},\mathbf{e}_{2},\mathbf{e}_{3})\equiv
(\mathbf{l},\mathbf{v},\mathbf{w})$, where $\mathbf{v}$ and
$\mathbf{w}$ are unit vectors perpendicular to each other and to the
vector $\mathbf{l}$, the gravitational radiation power at frequency
$\omega_n$ per solid angle $d\Omega$ is given by~\cite{Durrer}
\begin{equation}
  \label{P:dur} \frac{d P_n}{d
  \Omega}=\frac{G \omega_n^2}{\pi} [{\tau}^{*}_{pq}
  {\tau}_{pq}-\frac{1}{2}{\tau}^{*}_{qq}{\tau}_{pp}] \,,
\end{equation}
where $\tau^{pq}$ is the Fourier transform of $T^{\mu\nu}$ in the
corotating basis. Note that the indices $p,q$  in Eq.~(\ref{P:dur})
take only the values $2$ and $3$.

The energy-momentum $T^{\mu\nu}_{\rm str}$ of a string is given by
\begin{equation}
  \label{Tstr}
  T^{\mu\nu}_{\rm str}= \mu\int
  d\sigma\left(\dot{x}^{\mu}\dot{x}^{\nu}-x^{\prime\mu}x^{\prime\nu}\right)
  \delta^{(3)}\left(\mathbf{x}-\mathbf{x}(\sigma,t)\right) \,.
\end{equation}
The Fourier components $\tau_{pq}$  for a string moving according to the
system~(\ref{sol}) can be expressed in the corotating basis as
\begin{eqnarray}
  \label{taustr}
   \tau^{pq}_{\rm str}(\omega_n,\mathbf{l})&=&
  \frac{\mu}{\pi R}\int_{\sigma_1}^{\sigma_2} d\sigma
  \int_0^{\pi R}dt
  e^{i\omega_n(t-\mathbf{l}\mathbf{x})} \nonumber \\
  &\times& \left[ (\dot\mathbf{x}\mathbf{e}_p)(\dot\mathbf{x}\mathbf{e}_q)
   -(\mathbf{x}'\mathbf{e}_p)(\mathbf{x}'\mathbf{e}_q)  \right] \,.
\end{eqnarray}
Introducing the new variables  $\tau=t/R$ and $\zeta=\sigma/R$ results
in
\begin{eqnarray}
  \label{taustr1}
  \tau^{pq}_{\rm str}=
  -\mu R\int_{-\zeta_0}^{\zeta_0} \!\!\!d\zeta\,
  \cos^2\zeta\,\chi^{pq}
  +\mu R\int_{-\zeta_0}^{\zeta_0} \!\!\!d\zeta\,
  \sin^2\zeta\,\psi^{pq} \,,
\end{eqnarray}
where $\zeta_0=\arcsin(L/2R)$
and
\begin{eqnarray}
  \label{chipsi}
   \chi^{pq}&=&
  \frac{1}{2\pi}\int_{0}^{2\pi} \!\!\!  d\tau\,
  e^{2in(\tau-\sin\zeta\,\mathbf{n}\mathbf{y}(\tau))}\,
  (\mathbf{y}(\tau)\mathbf{e}_p)\,(\mathbf{y}(\tau)\mathbf{e}_q),\nonumber\\
 \psi^{pq}&=&
  \frac{1}{2\pi}\int_{0}^{2\pi}  \!\!\! d\tau\,
  e^{2in(\tau-\sin\zeta\,\mathbf{n}\mathbf{y(\tau)})}\,
  (\dot\mathbf{y}(\tau)\mathbf{e}_p)\,(\dot\mathbf{y}(\tau)\mathbf{e}_q)
  \,. \nonumber
\end{eqnarray}

The energy-momentum tensor $T^{ij}_{\rm mon}$ of two point-like
particles moving according to Eq.~(\ref{sol}) is given by
\begin{equation}
  \label{Tmon}
  T^{ij}_{\rm mon}(t,\mathbf{x})
  =m\gamma_0[\delta (\mathbf{x}-\mathbf{x}_1(t))+
  \delta (\mathbf{x}-\mathbf{x}_2(t))]
  \dot{x}_1^i(t)\dot{x}_1^j \,,
\end{equation}
where $\gamma_0=(1-(L/2R)^2)^{-1/2}$ is the Lorentz factor of the
point-like masses. The Fourier components of the energy-momentum
tensor in the corotating basis follow as
\begin{eqnarray}
  \label{taumon}
  \tau^{pq}_{\rm mon}&=&\gamma_0 m \left(\frac{L}{2R}\right)^2
  \frac{1}{2\pi} \int_0^{2\pi}\,d\tau\,
   \left[e^{2in(\tau-L/2R\,\mathbf{l}\mathbf{y}(\tau))}\right.\nonumber\\
     &+&\left.e^{2in(\tau+L/2R\,\mathbf{l}\mathbf{y}(\tau))}\right]
  (\dot\mathbf{y}(\tau)\mathbf{e}_p)\,(\dot\mathbf{y}(\tau)\mathbf{e}_q)
  \,.
\end{eqnarray}
Applying the relation
\begin{equation}
  \frac{\mu}{m}=\frac{2}{L}\gamma_0^2\left(\frac{L}{2R}\right)^2
   =\frac{\sin\zeta_0}{R\cos^2\zeta_0} \,,
\end{equation}
then gives
\begin{eqnarray}
  \label{taumon1}
  \tau^{pq}_{\rm mon}&=&R\mu \sin\zeta_0\,\cos\zeta_0
  \frac{1}{2\pi} \int_0^{2\pi}\,d\tau\,
   \left[e^{2il(\tau-L/2R\,\mathbf{n}\mathbf{y}(\tau))}\right.\nonumber\\
     &+&\left. e^{2il(\tau+L/2R\,\mathbf{n}\mathbf{y}(\tau))}\right]
    (\dot\mathbf{y}(\tau)\mathbf{e}_p)\,(\dot\mathbf{y}(\tau)\mathbf{e}_q) \,.
\end{eqnarray}

The Fourier transform of the total energy-momentum tensor
$\tau^{pq}_{\rm total}$ is the sum of $\tau_{\rm str}$ and
$\tau_{\rm mon}$. The final expression for $\tau^{pq}_{\rm total}$
is (see Appendix~\ref{App:EMT} for details of the calculation)
\begin{eqnarray}
  \label{tautotal}
   \tau^{pq}_{\rm total}&=&\frac{\mu R}{\sin\theta}\times\\
    &\times&\left[\int_0^{\zeta_0}
    \mathbf{R}_n^{pq}(\sin\theta,\,\zeta)\,d\zeta+
     \mathbf{S}^{pq}_n(\sin\theta,\,\zeta_0)\right] \,,\nonumber
\end{eqnarray}
where
\begin{eqnarray}
  \label{R}
   \mathbf{R}^{22}_n(x,z)&=&-\frac{2\cos^2\theta}{x}J_{2n}(2nx\sin z),\nonumber\\
   \mathbf{R}^{23}_n(x,z)&=& -2i\cos\theta J'_{2n}(2nx\sin z)\,\sin z ,\\
   \mathbf{R}^{33}_n(x,z)&=& \frac{2}{x}J_{2n}(2nx\sin z)\left[1+x^2 \cos(2z)\right],\nonumber
\end{eqnarray}
and
\begin{eqnarray}
  \label{S}
   \mathbf{S}^{22}_n(x,z)&=&
    -\frac{2(1-x^2)}{x}J_{2n}(2nx\sin z)\,\frac{\cos z}{\sin z},\nonumber\\
   \mathbf{S}^{23}_n(x,z)&=& -i\sqrt{1-x^2}J'_{2n}(2nx\sin z)\,\cos z,\\
   \mathbf{S}^{33}_n(x,z)&=&
    \frac{2}{x}J_{2n}(2nx\sin z)\,\frac{\cos z}{\sin z}
     \left[1-x^2\sin^2 z\right].\nonumber
\end{eqnarray}

Let us now consider the limit of heavy monopoles or short, light
strings, i.e., $\mu R/m\ll 1$ or $\zeta_0\ll 1$. Keeping only terms up
to first order in $\zeta_0$, the integrals in Eq.~(\ref{tautotal}) containing
$\mathbf{R}_l^{pq}$ can be neglected. This means that the contribution
of the string to the total energy-momentum tensor is negligible.
For $\mathbf{S}_l^{pq}$ one finds
\begin{eqnarray}
  \label{S1}
   \mathbf{S}_1^{22} &=& \mathbf{S}_1^{23}=
    \mathbf{S}_1^{33}\simeq x\zeta_0,\nonumber\\
    \mathbf{S}_l^{pq}&\simeq&0,\quad l\ge 2 \,.
\end{eqnarray}
Substituting Eq.~(\ref{S1}) into Eq.~(\ref{tautotal}),
the components of the energy-momentum are in the limit $\mu R/m\ll 1$
\begin{eqnarray}
  \label{totaltau1}
   \tau_{\rm total}^{22}&=&-\frac{(\mu R)^2}{m}\cos^2\theta,\nonumber \\
   \tau_{\rm total}^{23}&=&-i\frac{(\mu R)^2}{m}\cos\theta,\\
   \tau_{\rm total}^{33}&=&\frac{(\mu R)^2}{m}\,. \nonumber
\end{eqnarray}
This expression coincides with the one calculated using the quadrupole
approximation~\cite{Weinberg} for moving particles,
\begin{equation}
  \label{quadr}
   \tau^{pq}=-(\omega^2/2)\int\,d^3x\,(\mathbf{x}\mathbf{e}^p)
   (\mathbf{x}\mathbf{e}^q)\, T^{00}(\omega,\mathbf{x}) \,.
\end{equation}

The opposite case of light particles, $\mu R/m\gg 1$ or
$\zeta_0-\pi/2\ll 1$, is more complicated. For high frequencies, the
radiation is highly concentrated in the plane of rotation of the system.
For not too high mode numbers, $1\ll n\ll n_{\rm cr}$, where
$n_{\rm cr}$ is given by
\begin{equation}
  \label{cut-off}
   n_{\rm cr}= \left(\frac{\mu R}{m}\right)^{3/2} \,,
\end{equation}
the radiation rate of the whole system behaves as if there are no
monopoles at the ends of the string: Using the expression
for the Lorentz-factor in the ultra-relativistic limit,
\begin{equation}
  \label{Lorentz}
   \gamma_0=\left(\frac{\mu R}{m}\right)^{1/2},
\end{equation}
the cut-off mode number (\ref{cut-off}) can be rewritten as
\begin{equation}
  \label{cut-off1}
   n_{\rm cr}= \gamma_0^3 \,.
\end{equation}
In the range $1\ll n\ll n_{\rm cr}$, the gravitational
radiation is well approximated by (see Appendix \ref{App:HFA})
\begin{equation}
 \label{Pnth}
  P_n=\frac{C G\mu^2}{n} \,,
\end{equation}
where the coefficient $C$ is given by Eq.~(\ref{App:C}); its
numerical value is $C\simeq 5.77$. For $1\ll n\ll n_{\rm cr}$, the
gravitational radiation depends only on the string tension $\mu$ and
not on the monopole mass $m$. In the absence of monopoles, $m\to 0$,
the total radiated power diverges logarithmically. In this case,
the back reaction of the gravitational radiation on the string
should be taken into account. In contrast,  the gravitational spectrum
from cosmic strings with monopoles on their ends has a finite cut-off
frequency and, thus, the total emitted gravitational radiation is finite.

For very high mode numbers, $n\gg n_{\rm cr}$, the radiated
gravitational power depends on the ratio $\mu R/m$ as well as on $\mu$
and is given by (see Appendix \ref{App:HFA})
\begin{eqnarray}
  \label{Pvh}
    P_n &=& G\mu^2
     \sqrt{\frac{2}{\pi}} \left(\frac{m}{\mu R}\right)^{9/4}
     \sqrt{n}\,\exp\left[-\frac{4n}{3}\left(\frac{m}{\mu R}\right)^{3/2}\right]\nonumber\\
      &=& G\mu^2 \sqrt{\frac{2n}{\pi}}\,\gamma_0^{9/2}\,e^{-4n\gamma_0^3/3}\,.
\end{eqnarray}

\begin{figure}[t]
\includegraphics[width=0.48\textwidth]{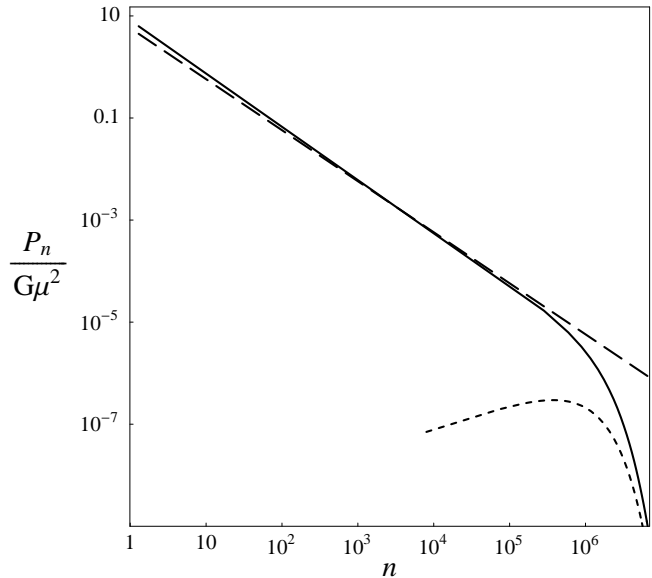}
\caption{\label{P_n} A log-log plot of the gravitational spectrum
$P_n/G\mu^2$ for $\gamma=100$ (solid line) is shown. The approximation
(\ref{Pnth}) is plotted by dashed line and high frequency approximation
(\ref{Pvh}) is plotted by dotted line.}
\end{figure}

An example of the gravitational spectrum for $\gamma_0=100$ is shown in
Fig.~\ref{P_n} together with the two asymptotic expressions, Eq.~(\ref{Pnth})
and Eq.~(\ref{Pvh}). For very small $n$, the discrete spectrum of the
radiated energy cannot be neglected and the low-frequency
approximation underestimates the gravitational radiation slightly,
while in the range $1\ll n\lsim\gamma_0^3$ the approximation
describes well the exact result. For $n\gsim\gamma_0^3$, the
high-frequencies limit approximates well Eq.~(\ref{Pvh}).

\begin{figure}[t]
\includegraphics[width=0.48\textwidth]{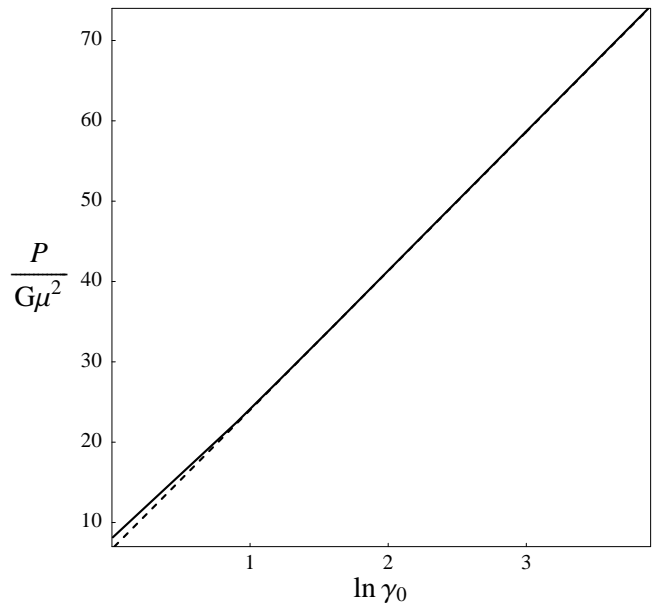}
\caption{\label{P} the total gravitational
radiation $P$ (solid line) and the approximation (\ref{Ptot})
(dashed line) is shown as a function of $\ln\gamma_0$.}
\end{figure}

For moderate values of $n_{\rm cr}$, the total gravitational radiation
power $P$ can be calculated by summing up the radiation power at
different modes $P_n$, while for larger values of $n_{\rm cr}$
a semi-analytical formula for $P$ is useful. With $n_{cr}\propto
\gamma_0^3$, we obtain as total gravitational power
\begin{equation}
  \label{Ptot}
   P\simeq G\mu^2(3 C \ln\gamma_0+B) \,.
\end{equation}
The numerical coefficient $B\simeq 6.8$ accounts for the deviation of
the low-frequency limit Eq.~(\ref{Pnth}) from  $P_n$ for $n\to 1$.
In Fig.~\ref{P}, the total gravitational
radiation $P$ (solid line) and the approximation (\ref{Ptot})
(dashed line) is shown as a function of $\ln\gamma_0$.

At the end of this section, we want to compare our results for the
gravitational radiation from a rotating monopole-string system with
those for an oscillating monopole-string system derived in
Ref.~\cite{MarVil}. There are two main difference between the rotating
and the oscillating system: Firstly, the Lorentz factor of the
monopoles depends differently on the dimensionless quantity
$\mu R/m$. While $\gamma_0=(\mu R/m)^{1/2}$ for the
rotating rod, $\tilde{\gamma}_0=\mu R/m$ for the oscillating
solution~\cite{MarVil}.  Secondly, the critical mode number
$n_{\rm cr}$ is for the oscillating solution given by
$n_{\rm cr}=\tilde\gamma_0^2\sim (\mu R/m)^2$ compared to
$n_{\rm cr}=\gamma_0^3\sim (\mu R/m)^{3/2}$ for the rotating rod.

For modes below the critical number $n_{\rm cr}$ of both solutions, the
spectrum of emitted gravitational waves is surprisingly similar:
$P_n/(G\mu^2)\simeq 4/n$ for the oscillating solution and
$P_n/(G\mu^2)\simeq 5.77/n$ for the rotating rod solution. The
difference becomes more pronounced only for frequencies above
$n_{\rm cr}$. While the high-frequency limit of the rotating rod
solution is well-behaved and modes with $n \gg n_{\rm cr}$ are
exponentially suppressed, the emitted power behaves as $P_n\propto
n^{-2}$ for the oscillating solution. This $1/n^2$ behavior is caused
by the discontinuity of the monopole acceleration~\cite{MarVil} and is
probably unphysical. In the generic case of a superposition of an
oscillating and rotating monopole-string system we expect therefore an
exponential decay of the radiated power similar to the Eq.~(\ref{Pvh}).
We estimate therefore the emitted gravitational radiation in the
generic case as
\begin{equation}
\label{P_nest}
 P_n/(G\mu^2)
  =\left\{\begin{array}{lcl} 5/n,& & n\alt n_{\rm cr}, \\
 0,& & n\agt n_{\rm cr} \\
\end{array} \right.,
\end{equation}
where $n_{\rm cr}$ is given by (\ref{cut-off}). The total gravitational
power in the generic case follows as
\begin{equation}
  \label{Ptotest}
    P \sim 7G\mu^2\ln\left(\frac{\mu R}{m}\right)\,.
\end{equation}

\section{Detection of the gravitational wave signal}
\label{network}

In this section we study the possibility to detect the gravitational
radiation from oscillating and rotating monopole-string systems.
In the first part, we consider the gravitational radiation from a
network of monopoles and antimonopoles connected by $N$ strings.
The evolution of such a network can be analyzed in two limiting
cases. If monopoles are light enough, then the cosmic string network
does not feel the presence of monopoles. Thus we obtain the ``normal''
scaling solution for the network evolution where the typical
length $\xi$ of closed strings is
\begin{equation}
  \label{length}
  \xi\sim \Gamma G\mu t \,.
\end{equation}
Here, $t$ is the cosmological time and the dimensionless coefficient
$\Gamma\sim 50$ determines the gravitational radiation from strings.
In the opposite limit, we may assume that
the gravitational radiation from monopoles is the dominant energy loss
mechanism. Then we find, following Ref.~\cite{VacVil}, another scaling
solution for the typical distance $d$ between monopoles with
\begin{equation}
  \label{length2}
  d\sim \tilde\Gamma G\mu t \,,
\end{equation}
where $\gamma$ is the typical Lorentz-factor of the monopoles and
$\tilde\Gamma\sim 10\ln\gamma$. Apart from an numerically not too
important factor $5/\ln\gamma$, the two limits give the same
characteristic length $\xi$ for the network
\footnote{Actually, the additional limit
$\tilde\Gamma\le \Gamma$ should be imposed because
for very high Lorentz-factors $\gamma$ the
monopoles lose their energy to the gravitational
radiation less efficiently than given by (\ref{Ptotest}). This is
because for very high $\gamma$ the monopoles does not
accelerated linearly like $a\sim \mu/m$
but are involved in the general string motion. Therefore
we can estimate the typical scale $\xi$ of the necklace network
by the ``usual'' formula (\ref{length}), where $\xi$ is the closed
loop length or the distance between the monopoles.}.

The presence of monopoles (and antimonopoles)
on the strings changes the GWBs from cusps of cosmic strings.
Massive particles living on the
string affect the cusp formation which happens only in the case of
``ordinary'' Nambu-Goto strings. Adding monopoles makes strings
heavier and smoothes-out cusps, similar to the presence of superconducting
currents on the string. The consequence of smoothed cusps is a cut-off
in the emitted gravitational radiation~\cite{BabDok}. Therefore we assume that the
presence of monopoles on the strings leads to a cut-off  given by
Eq.~(\ref{cut-off}),
\begin{equation}
  \label{cut-offw}
   \omega_{\rm cr}\sim \frac{\gamma_0^3}{\xi} \,.
\end{equation}

In Ref.~\cite{DamVil1}, Damour and Vilenkin
argued that the gravitational bursts from
ordinary string cusps can be detected by the gravitational wave
detectors like LIGO, VIRGO and especially LISA.
How is this analysis changed by the presence of monopoles on the
strings? The main difference of the radiation of a
string-monopole network from the radiation of an ordinary string network
is the existence of a cut-off frequency, such that the signal from
cusps with frequencies $\omega\agt\omega_{cr}$ is highly
suppressed. 

The cut-off frequency of the arriving signal is
\begin{equation}
  \label{cr}
   \omega_{\rm cr}=\gamma_0^3\,\left(\Gamma G\mu t_0 \right)^{-1}\times\left\{\begin{array}{lcl}
     (1+z)^{1/2}, & & z\alt z_{\rm eq}, \\
     (1+z)(1+z_{\rm eq})^{-1/2}
      & &  z\agt z_{\rm eq}, \\
    \end{array} \right. \,,
\end{equation}
where at redshift $z_{\rm eq}$ matter and radiation densities are equal.

The typical length of loops decreases with $z$ and
consequently the cut-off frequency increases with red-shift.
Using Eq.~(\ref{length}) and $R\sim \xi$, the behavior
of the Lorentz-factor as function of the redshift follows as
\begin{equation}
  \label{Lorentz_cosm}
   \gamma_{0}= \left(\Gamma G\mu \right)^{1/2}\left(\frac{\mu t_0}{m}\right)^{1/2}\times
    \left\{\begin{array}{lcl}
     (1+z)^{-3/4}, & & z\alt z_{\rm eq}, \\
      \frac{(1+z_{\rm eq})^{1/4}}{1+z},  & &  z\agt z_{\rm eq}, \\
    \end{array} \right. \,.
\end{equation}
Hence, the cut-off frequency is given by
\begin{eqnarray}
  \label{cr1}
   \omega_{\rm cr}&=&
   (\Gamma G\mu)^{1/2}\left(\frac{\mu t_0}{m}\right)^{3/2}t_0^{-1}\times\\
    &\times&\left\{\begin{array}{lcl}
     (1+z)^{-7/4}, & & z\alt z_{\rm eq}, \\
      (1+z_{\rm eq})^{1/4}(1+z)^{-2}, & &  z\agt z_{\rm eq}. \nonumber\\
    \end{array} \right. \,.
\end{eqnarray}
The analysis given by Damour and Vilenkin \cite{DamVil1}
leads to the following result: for a given frequency $\omega$ and
for a given string parameter $\mu$ (or, strictly speaking, for a
given parameter $\Gamma G \mu$) the amplitude of incoming signals
$h$ from cusps in the universe decrease with $z$, but the number
of such signals per unit time $\dot{N}(\mu,\omega)$ increases with
$z$. Thus, we should find the compromise between the rate of
signals and their amplitude. For a given $\dot{N}$ one can find
the minimal amplitude of incoming signals $h(\mu,\dot{N},\omega)$
and corresponding maximal redshift $z_m(\mu,\dot{N},\omega)$.

How does the existence of the cut-off frequency (\ref{cr}) modify
these arguments? The presence of monopoles may effectively lower the
amplitude of the signal from cusp.
Let us estimate the minimal mass of monopoles that may affect
the amplitude $h$. There are three regimes of behavior of
$z_m$ in dependence of the function $y(\mu,\dot{N},\omega)$,
\begin{equation}
\label{y} y(\mu,\dot{N},\omega)=10^{-2}(\dot{N}/c)t_0^{5/3}
(\Gamma G\mu)^{8/3} \omega^{2/3},
\end{equation}
where $c$ is the the average number of cusps per loop period
(usually it is taken that $c\sim 0.1-1$). The maximal redshift $z_m(y)$
can be expressed as~\cite{DamVil1}
\begin{equation}
  \label{zm}
   z_m(y)=\left\{\begin{array}{lcl} y^{1/3}, & & y\alt1,\\
   y^{6/11}, & & 1\alt y\alt y_{\rm eq},\\
   (y_{\rm eq}\,y)^{3/11}, & & y_{\rm eq}\alt y,\\
 \end{array}
\right. \,,
\end{equation}
where $y_{\rm eq}=z_{\rm eq}^{11/6}$. Inserting (\ref{zm}) in (\ref{cr})
we can find from the equation $\omega_{\rm cr}=\omega$ the critical
value $\mu t_0/m$ for which the monopoles suppress the  amplitude of
the gravitational signal from the network,
\begin{equation}
  \label{ecr}
    \frac{\mu t_0}{m}=\frac{(\omega t_0)^{2/3}}{(\Gamma G\mu)^{1/3}}
      \times \left\{\begin{array}{lcl}
     1,& & y\alt 1,\\
     y^{7/11},& & 1\alt y\alt y_{\rm eq},\\
     y_{\rm eq}^{3/11}\,y^{4/11}, & & y_{\rm eq}\alt y.\\
   \end{array}
\right.
\end{equation}
Let us fix the rate of observable GWBs $\dot{N}/c\sim 1$ ${\rm yr}^{-1}$,
the frequency
$\omega\sim 10^2$ (preferable for detecting GWBs with LIGO/VIRGO)
and find $\mu t_0/m$ as a function of
$G\mu$. From (\ref{ecr}) we obtain
\begin{equation}
  \label{ecr1}
    \frac{\mu t_0}{m}=\left\{\begin{array}{lcl}
     10^{12}\,(G\mu)^{-1/3},\,& & G\mu\alt 10^{-10},\\
     10^{29}\,(G\mu)^{15/11},\,& & 10^{-10}\alt G\mu\alt 10^{-7},\\
     10^{24}\,(G\mu)^{7/11},\,& & G\mu\agt10^{-7}.\\
   \end{array}
\right.
\end{equation}
For instance, for strings with $G\mu\sim 10^{-13}$
we have $\mu t_0/m\sim 10^{16}$, while for grand
unified theory strings with
$G\mu\sim 10^{-6}$  we find $\mu t_0/m\sim 10^{20}$.
Thus the monopoles should have masses well beyond the Planck mass, which,
of course, can not be realized in nature.
Therefore we conclude that the presence of monopoles
on the string network does not affect the observation of gravitational
wave burst from a cosmic string network.

In the end of this section, let us consider the gravitational
radiation background from  hybrid topological defects,
formed in the sequence of phase transitions with $N=1$, (\ref{G2}).
The evolution of single monopole-antimonopole pairs connected by one
string is very different from the one considered above. This system 
has a lifetime much less than the Hubble time, i.e. does not survive
till now. The efficient emission of gravitational waves depends
crucially when the phase transition producing monopoles happens.
If monopoles are formed after inflation and later get connected by strings 
(e.g. as in the Langacker-Pi model~\cite{lp}), the monopole-string
system loses most of its energy due to friction with the surrounding
plasma and a negligible gravitational wave background  results.
However, in another possible scenario proposed  in Ref.~\cite{MarVil1}
traces of topological defects might be left in the form of
gravitational waves. In this scenario, monopoles are formed during inflation, 
but are not swept away completely. The evolution of the monopole-string
system is even during the period of large friction similar to the 
evolution of ``ordinary'' strings without monopoles, because of 
the large monopoles separation. Therefore, the monopole-antimonopole pairs
connected by strings survive the period of large dumping 
forces. After that, the string-monopole systems begin to move 
relativistically and radiate most of their energy into gravitational
waves.
Martin and Vilenkin \cite{MarVil1}
estimated the gravitational wave background from such hybrid
defects using for the calculations the gravitational spectrum of the
oscillating solution.  The results of Ref.~\cite{MarVil1}
do not change strongly for a generic moving monopole-string system,
because of two reasons: (i) at low
frequencies the spectra are almost the same except an unimportant
factor $1.4$ and (ii) the different dependence of the cut-off frequency
on the parameter $\mu R/m$ enters in the formula for the total power emitted
only logarithmically. Thus the conclusion of Ref.~\cite{MarVil1} that
the gravitational wave background might be observable by
advanced interferometers does not change.

\section{Conclusion}
\label{Conclusion}

We have studied the radiation of gravitational waves by a
``rotating rod'', i.e. a rotating monopole-antimonopole pair
connected by a cosmic string. We have found that the dimensionless
parameter $\mu R/m$ determines the main features of the emitted
spectrum: For not too high mode numbers,
$n \ll n_{\rm cr}=(\mu R/m)^{3/2}$, the gravitational wave spectrum
can be approximated by  $P_n\approx 5.77 G\mu^2/n$, while the
gravitational radiation is exponentially suppressed  for $n\gg n_{\rm cr}$.
The total gravitational radiation from the rotating rod is given by
$P\simeq G\mu^2(17.3\ln\gamma +6.8)$.

The total power emitted and also the spectrum at not too
high frequencies  found by us agrees approximately with the spectrum
of an oscillating monopole-antimonopole pair connected by a string
from Ref.~\cite{MarVil}. Only in the high
frequency limit differences appear: While the gravitational radiation is
as expected on physical grounds exponentially suppressed for the
rotating solution, it decays only as $1/n^2$ for the oscillating
solution. Therefore, we have concluded  that the spectrum found by us applies
approximately also to the generic case of a superposition of a
rotational and oscillatory motion of the monopole-string system.
Moreover,
studies about the potential of interferometers to observe gravitational
wave emitted by topological defect networks that were based on the
oscillating solution are valid also for more general motions of the
monopole-string systems. We have confirmed that advanced gravitational
wave detectors like LIGO, VIRGO and LISA have the potential to detect
a signal topological hybrid defects for string tensions as small as
$G\mu\sim 10^{-13}$.

\begin{acknowledgments}
In Munich, this work was supported by an Emmy-Noether grant of the
Deutsche Forschungsgemeinschaft and in Moscow by the Russian
Foundation for Basic Research, grants 03-02-16436 and 04-02-16757.
\end{acknowledgments}


\appendix
\section{Energy-momentum tensor}
\label{App:EMT}

In this Appendix, we present the details of our calculatation of the
energy-momentum tensor of a rotating rod. The considered system is
axially symmetric and, therefore, it is sufficient to calculate
the radiation only in the plane perpendicular to the plane of
rotation. For definiteness, let us choose the following corotating
basis,
\begin{eqnarray}
  \label{App:basis}
  \mathbf{l}&=&\{\sin\theta,0,\cos\theta\},\nonumber\\
  \mathbf{v}&=&\{\cos\theta,0,-\sin\theta\}, \\
  \mathbf{w}&=&\{0,1,0\} \,.\nonumber
\end{eqnarray}
Substituting (\ref{App:basis}) into (\ref{chipsi}), one obtains
\begin{eqnarray}
  \label{App:chi}
  \chi^{22}&=&\frac{1}{2}\cos^2\theta\left[J^{'}_{2n-1}(2nr)-
    J^{'}_{2n+1}(2nr)\right],\nonumber\\
  \chi^{23}&=&\frac{i}{2}\cos\theta\left[J^{'}_{2n-1}(2nr)+
    J^{'}_{2n-1}(2nr)\right],\\
 \chi^{33}&=&-\frac{1}{2}\left[J^{'}_{2n-1}(2nr)-
    J^{'}_{2n+1}(2nr)\right]-J_{2n}(2nr) \,,\nonumber
\end{eqnarray}
and
\begin{eqnarray}
  \label{App:psi}
  \psi^{22}&=& -\frac{1}{2}\cos^2\theta\times\nonumber\\
   &\times&\left[\frac{2n-1}{2nr}J_{2n-1}(2nr)+
    \frac{2n+1}{2nr}J_{2n+1}(2nr)\right],\nonumber\\
  \psi^{23}&=&-\frac{i}{2}\cos\theta\times\\
    &\times&\left[\frac{2n-1}{2nr}J_{2n-1}(2nr)-
    \frac{2n+1}{2nr}J_{2n+1}(2nr)\right],\nonumber\\
  \psi^{33}&=&\frac{1}{2}\left[\frac{2n-1}{2nr}J_{2n-1}(2nr)
    +\frac{2n+1}{2nr}J_{2n+1}(2nr)\right]\nonumber\\
    &-&J_{2n}(2nr)\,,\nonumber
\end{eqnarray}
where $J_n(z)$ is $n$-th Bessel function and
$r=|\sin\zeta\,\sin\theta|$. Inserting (\ref{App:chi}) and
(\ref{App:psi}) into (\ref{taustr1}), we find as components of the
energy-momentum tensor of the string
\begin{eqnarray}
  \label{App:taustr}
  \tau^{22}_{\rm str}&=&-2\mu R\,\frac{\cos^2\theta}{\sin^2\theta}\nonumber\\
   &\times&\left\{\int_0^{\zeta_0} J_{2n}(2nr)d\zeta
    -\frac{\sin\theta}{2n}J'_{2n}(2nr_0)\cos\zeta_0 \right\},\nonumber\\
  \tau^{23}_{\rm str}&=&-\mu R\,\frac{i\cos\theta}{\sin\theta}\nonumber\\
   &\times&\left\{\int_0^{\zeta_0} \left[J_{2n-1}(2nr)-J_{2n+1}(2nr)\right]d\zeta
   \right.\nonumber\\
   &-&\left.\frac{1}{n\,\sin\theta\,\sin\zeta_0}J_{2n}(2nr_0)\cos\zeta_0 \right\},\\
   \tau^{33}_{\rm str}&=&2\mu R\,\frac{1}{\sin^2\theta}\nonumber\\
      &\times&\left\{\int_0^{\zeta_0} \left[J_{2n}(2nr)
        +\sin^2\theta\,\cos(2\zeta)\,J_{2n}(2nr)\right]\,d\zeta\right. \nonumber\\
      &-&\left.\frac{\sin\theta}{2n}J'_{2n}(2nr_0)\cos\zeta_0 \right\}.\nonumber
\end{eqnarray}

Let us turn now to the calculation of the energy-momentum tensor
of the monopoles. Using the fact that the expression
(\ref{taumon1}) is very similar to the expression for
$\psi^{pq}$ in (\ref{chipsi}), one easily obtains
\begin{eqnarray}
  \label{App:taumon}
   \tau^{22}_{\rm mon}&=&-\mu R\cos\zeta_0\,\frac{\cos^2\theta}{\sin\theta}\nonumber\\
     &\times&\left[\frac{2n-1}{2l}J_{2n-1}(2nr_0)+
      \frac{2n+1}{2n}J_{2n+1}(2nr_0)\right],\nonumber\\
   \tau^{23}_{\rm mon}&=&-\mu R\cos\zeta_0\,\frac{i\cos\theta}{\sin\theta}\\
     &\times& \left[\frac{2n-1}{2n}J_{2n-1}(2nr_0)-
      \frac{2n+1}{2n}J_{2n+1}(2nr_0)\right],\nonumber\\
   \tau^{33}_{\rm mon}&=&\mu R\,\cos\zeta_0\frac{1}{\sin\theta}\nonumber\\
     &\times&\left[\frac{2n-1}{2n}J_{2n-1}(2nr_0)+
      \frac{2n+1}{2n}J_{2n+1}(2nr_0)\right]\nonumber\\
      &&-\mu R\sin(2\zeta_0)J_{2n}(2nr_0) \,.\nonumber
\end{eqnarray}
Summing then (\ref{App:taustr}) and (\ref{App:taumon}), we finally obtain
the Fourier components of energy-momentum of the whole system.

\section{High-frequency behavior}
\label{App:HFA}

Throughout this Appendix we assume that the monopoles
are very light, i.e. $R\mu/m\gg 1$.
The qualitative behavior of the  radiation emitted by a rotating string
at high frequencies can be derived using the following asymptotic
relations for Bessel functions of large order (cf.~Ref.~\cite{AS}),
\begin{equation}
  \label{App:Bessel_asy}
   J_n(nx)\propto \left\{\begin{array}{lcl}
    n^{-1/3},& & n\ll n_{\rm cr}, \\
    \exp(-n/n_{\rm cr})& & n\gg n_{\rm cr}, \\
                        \end{array}
                        \right.
\end{equation}
and
\begin{equation}
  \label{App:Bessel_asy1}
   J'_n(nx)\propto \left\{\begin{array}{lcl}
    n^{-2/3},& & n\ll n_{\rm cr}, \\
    \exp(-n/n_{\rm cr})& & n\gg n_{\rm cr}, \\
                        \end{array}
                        \right.
\end{equation}
where $n_{\rm cr}$ is the cut-off frequency given by
\begin{equation}
  \label{App:cut-off}
   n_{\rm cr}\simeq \frac{3}{(\pi/2-\zeta_0)^3}=
   3\left(\frac{\mu R}{m}\right)^{3/2} \,.
\end{equation}
Using the relations (\ref{App:Bessel_asy}) and
(\ref{App:Bessel_asy1}), from (\ref{tautotal}) follows that
the radiated power $P_n$ behaves for $1\ll n\ll n_{cr}$ as
\begin{equation}
 \label{App:P_simple}
  P_n\propto n^{-1} \,,
\end{equation}
which was pointed by Martin and Vilenkin~\cite{MarVil}. If there are
no point-like masses on the ends of a string, then the total
radiated power diverges because of presents of string points, permanently
moving with the speed of light. The presence of particles
of any non-zero mass at the end-points removes such a divergence.
For $n\gg n_{cr}$, the radiated gravitational power falls
exponentially (as can be seen from (\ref{App:Bessel_asy}) and
(\ref{App:Bessel_asy1}), thus making the sum over all modes finite.

Let us calculate now the high-frequency behavior of
the gravitational radiation more precisely.
For not very high frequencies, $1\ll n\ll n_{\rm cr}$, the contribution
from monopoles to the gravitational radiation may be neglected. In
this case we can simply set $\zeta_0=\pi/2$ in the expression for the
components of the energy-momentum tensor, Eq.~(\ref{tautotal}).
Then $\mathbf{S}^{pq}_n=0$ and, moreover, the integrals
in Eq.~(\ref{tautotal}) can be integrated in closed form.
Using the properties of integrals involving Bessel functions, we
obtain
\begin{eqnarray}
  \label{App:totaltau1}
   \tau^{22}_{\rm total}&=&-\pi\mu R\,\frac{\cos^2\theta}{\sin^2\theta}
    J_n^2(n\sin\theta),\nonumber\\
   \tau^{23}_{\rm total}&=&-i\pi\mu R\,\frac{\cos\theta}{\sin\theta}
    J_n(n\sin\theta)\,J'_n(n\sin\theta),\\
   \tau^{33}_{\rm total}&=&\pi\mu R\,
    J^{'2}_n(n\sin\theta) \,.\nonumber
\end{eqnarray}
Substituting these expressions into Eq.~(\ref{P:dur}) results in
\begin{eqnarray}
  \label{App:P:nvhf}
   \frac{d P_n}{d\Omega} &=& 2\pi\,G\mu^2\,n^2
   \left[\frac{\cos^4\theta}{\sin^4\theta}\,J_n^4(n\sin\theta)\right.\\
    &+&\left. 6\frac{\cos^2\theta}{\sin^2\theta}\,J_n^2(n\sin\theta)J_n^{'2}(n\sin\theta)
   +J_n^{'4}(n\sin\theta)\right] \,,\nonumber
\end{eqnarray}
and, introducing the variable $x=\sin\theta$,
\begin{eqnarray}
  \label{App:P:Vil}
    P_n &=& 8\pi^2 G\mu^2 n^2 \int_0^1 \frac{x\,dx}{\sqrt{1-x^2}}
    \left[\frac{(1-x^2)^2}{x^4}\,J_n^4(nx)\right. \nonumber\\
     &&\left. + 6\,\frac{1-x^2}{x^2} J_n^2(nx) J_n^{'2}(nx)
      +J_n^{'4}(nx)\right] \,.
\end{eqnarray}
This expression coincides with the result of Martin and Vilenkin for
a rotating string without monopoles~\cite{MarVil}.
To calculate the integral in (\ref{App:P:Vil}) we use further
the following approximations for Bessel functions and their
derivatives,
\begin{eqnarray}
  \label{App:Jappr}
   J_n(nx)&\simeq& \left(\frac{2}{n}\right)^{1/3}\,
    {\rm Ai}\left[2^{1/3}n^{2/3}(1-x)\right],\nonumber\\
    J'_n(nx)&\simeq& -\left(\frac{2}{n}\right)^{2/3}\,
    {\rm Ai'}\left[2^{1/3}n^{2/3}(1-x)\right] \,,
\end{eqnarray}
which is valid for $n\gg 1$. Here, ${\rm Ai(z)}$ is the Airy function.

Substituting (\ref{App:Jappr}) into (\ref{App:P:Vil}) and then
changing the variables to $u=2^{1/3}n^{2/3}(1-x)$, we obtain after some
algebra
\begin{equation}
 \label{App:P:C}
   P_n\simeq\frac{C G\mu^2}{n} \,,
\end{equation}
where the numerical factor $C$ is given by
\begin{eqnarray}
  \label{App:C}
   C &=& 32 \pi^2\int_0^\infty
    \frac{du}{\sqrt u}\left[u^2\,{\rm Ai}^4(u)\right.\nonumber\\
    &+&\left.6u{\rm Ai}^2(u){\rm Ai'}^2(u)+{\rm Ai'}^4(u)\right] \,.
\end{eqnarray}
(In the last integral we changed the limit $2^{1/3}n^{2/3}$
to $\infty$.) The numerical value of $C$ is $C\simeq 5.77$.

For $n\gg n_{\rm cr}$ the Airy functions in
(\ref{App:Jappr}) may be approximated
by exponentials resulting in the following
approximated expressions for Bessel function
and its derivative
\begin{eqnarray}
  \label{App:Jappr1}
   J_{2n}(2nz) &\simeq& \frac{1}{2\sqrt{\pi n}}\,
    \frac{e^{-2n(1-z^2)^{3/2}/3}}{(1-z^2)^{1/4}},\nonumber\\
   J'_{2n}(2nz) &\simeq& \frac{(1-z^2)^{1/4}}{2\sqrt{\pi n}}\,
    e^{-2n(1-z^2)^{3/2}/3} \,.
\end{eqnarray}
Substituting (\ref{App:Jappr1}) into (\ref{tautotal}), one finds
that the $\mathbf{S}$ terms are enhanced by the factor
$n(1-x^2\,\sin^2\zeta_0)^{3/2}\gg 1$ compared to the
$\mathbf{R}$ terms. Therefore, the $\mathbf{R}$ terms can be neglected.
Further, we introduce as a new variable $y=2n(1-x)\cos\zeta_0$ and
notice that the components of the energy-momentum tensor
(\ref{tautotal}) are proportional to the same exponential, but with
prefactors proportional to $n^{-3/2}$, $n^{-1}$ and
$n^{-1/2}$ for $\tau^{22}$, $\tau^{23}$, $\tau^{33}$, respectively.
Thus the leading term in the expression for the radiated power is the
$\tau^{33}$ term. Keeping only this component of the energy-momentum
tensor, expanding the argument of the exponent around $y=0$ and taking
the other slower varying terms as being constant, we obtain
\begin{eqnarray}
  \label{App:Pvh}
    P_n &\simeq& G\mu^2\cos^{9/2}\zeta_0\frac{\sqrt{2n}}{\pi}\times\nonumber\\
         &\times&  e^{-(4/3)n\cos^3\zeta_0}\int_0^{4n\cos\zeta_0}y^{-1/2}\,dy\,e^{-y}\\
    &\simeq&  G\mu^2 \left(\frac{m}{\mu R}\right)^{9/4} \frac{\sqrt{2n}}{\pi}\times\nonumber\\
    &\times&e^{-(4/3)n (m/\mu R)^{3/2}}\int_0^{\infty}y^{-1/2}\,dy\,e^{-y}\nonumber \\
    &=& G\mu^2 \sqrt{\frac{2}{\pi}} \left(\frac{m}{\mu R}\right)^{9/4}
     \sqrt{n}\,\exp\left[-\frac{4n}{3}\left(\frac{m}{\mu
         R}\right)^{3/2}\right] \,.\nonumber
\end{eqnarray}
As expected, the radiation rate is exponentially surpressed for $n\to\infty$.

\end{document}